# The twin paradox and the principle of relativity

Øyvind Grøn

Oslo University College, Faculty of Engineering, P. O. Box 4 St. Olavs Plass, N-0130 Oslo, Norway

**Abstract**

In the standard formulation of the twin paradox an "accelerated" twin considers himself as at rest and his brother as moving. Hence, when formulating the twin paradox, one uses the general principle of relativity, i.e. that accelerated and rotational motion is relative. The significance of perfect inertial dragging for the validity of the principle of relativity is made clear. Three new results are reviewed in the discussion: 1. A cosmic time effect which cannot be reduced to the gravitational or the kinematical time dilation; 2. Perfect dragging in an exact solution of Einstein's field equations describing flat spacetime inside a shell with Kerr spacetime outside it; 3. An extended model of Minkowski spacetime in order to avoid introducing absolute acceleration and rotation through the asymptotic emptiness of the Kerr spacetime.

1. Introduction

Two twins A and B meet at an event P1, move away from each other, and then meet again at a later event P2. The twin A considers himself as at rest and predicts that B is younger than himself at P2 due to the relativistic time dilation. But according to the principle of relativity B can consider himself as at rest and A as travelling, and he then predicts that A is younger when they meet at P2. The twin paradox is these contradicting predictions.

One may get rid of the twin paradox at once by noting that in order to be able to meet, depart and meet again, at least one of the twins must accelerate. And within the special theory of relativity the principle of relativity is not valid for accelerated motion. Acceleration is absolute. Hence, at least one of the twins is not allowed to consider himself as at rest. The twin with the greatest average velocity between the events P1 and P2 is youngest when the twins meet at P2.

However, in the present article I will use the twin paradox as a pedagogical entrance to the general theory of relativity. And the above resolution of the twin paradox is contrary to the spirit of the general theory of relativity. When Einstein presented this theory in 1916 [1], he wrote in the Introduction that the special theory of relativity contains a special principle of relativity, where the word "special" is meant to say that the principle is restricted to motion of uniform translation.

The second paragraph of Einstein's great 1916-article is titled: "The need for an extension of the postulate of relativity". He starts by writing that the restriction of the postulate of relativity to uniform translational motion is an inherent epistemological defect. Then he writes: "*The laws of physics must be of such a nature that they apply to systems of reference in any kind of motion*. Along this road we arrive at an extension of the postulate of relativity".

Furthermore Einstein makes use of the principle of equivalence, according to which the physical effects of inertial forces in an accelerated reference frame K' are equivalent to the effects of the gravitational force in a frame K at rest on the surface of a massive body. Einstein asks: Can an observer at rest in K' perform any experiment that proves to him that he is "really" in an accelerated system of reference? He says that this is not possible due to the principle of equivalence. Then he states: "Therefore, from the physical standpoint, the assumption readily suggests itself that the systems K and K' may both with equal right be looked upon as "stationary"".

Accepting this both twins have the right to consider themselves as at rest. We then have a much deeper and interesting twin paradox than the one restricted to the special theory of relativity which could be discarded at once due to the absolute character of accelerated motion in this theory.

In this article I will show how a discussion of the twin paradox within the conceptual frame of the general theory of relativity gives us the opportunity to discuss some of the most fundamental properties of the theory, for example: Does the theory imply the validity of the principle of relativity for accelerated and rotational motion? And what happens to the rate of time? Which twin ages fastest?

Two main points in the present article are the connection between the twin paradox and the principle of relativity, and the importance of the inertial dragging effect for the validity of the principle of relativity. I will also argue that the Kerr (and Schwarzschild) spacetime as usually understood with a globally empty remote Minkowski spacetime, is in conflict with the principle of relativity, and hence that an extended model of the Minkowski spacetime is needed in order that the principle of relativity shall be contained in the general theory of relativity.

## 2. The twin paradox and the principle of relativity

I will first consider the usual version of the twin paradox [2,3] in the Minkowski spacetime: Imagine that twin A remains at rest on the Earth and twin B travels with velocity *v = 0,8c* to the nearest star Alpha Proxima 4 light years from the Earth and back. According to A this will take

$$t_A(A) = 2L_0 / v = 10 \; years \tag{1}$$

which means that A is 10 years older at the reunion. A would predict that the twin B is

$$t_A(B) = \sqrt{1 - v^2/c^2} \; t_A(A) = 6 \; years \tag{2}$$

older at the reunion. *But according to the principle of relativity B could consider himself as at rest and A as moving.* According to the special theory of relativity B would then predict that he is 10 years older and A is 6 years older at the reunion. The contradiction between these predictions is the twin paradox.

The principle of relativity is essential for the formulation of the twin paradox. There would be no paradox if not both A and B could consider themselves as at rest. In the special theory of relativity non-accelerated motion is relative. This special principle of relativity is, however, not enough to formulate the twin paradox. In order that the twins shall be able to travel away from each other and reunite again, at least one of them must accelerate during the departure. So the general principle of relativity encompassing accelerated and rotating motion is needed.

There is still no general agreement on whether the general theory of relativity implies the validity of the general principle of relativity. This will be discussed later on. Let us for the moment assume that the general principle of relativity is valid, so that we have a twin paradox. The standard solution of the paradox requires some preparations.

When twin B arrives at Alpha Proxima he turns on the rocket, stops and then immediately accelerates towards his brother. As we saw, the general theory of relativity, i.e. the general principle of relativity, was needed in order to formulate the twin paradox. It should then not be too surprising that we need the general theory of relativity also to solve the paradox.

The general theory of relativity is based on the local (i.e. the region in spacetime is so small that tidal forces cannot be measured) validity of the special theory of relativity together with the principle of equivalence. According to the principle of equivalence the physical effects of an "artificial field of gravity" in an accelerated or rotating frame of reference are equivalent to the physical effects of a permanent field of gravity caused by a mass distribution.

Einstein deduced in 1911 the effect of a gravitational field upon the frequency of light. The point of departure was the Doppler effect: The frequency of a light signal gets an increase, i.e. the light gets a blue shift, if the light source moves towards the observer, and a red shift if the source moves away from the observer.

He then wrote about emission of light in an accelerated reference frame, for example in an accelerated rocket. First the situation is considered by an observer at rest. Light is emitted from the front end towards a detector further backwards. While the light moves from the emitter to the detector the rocket gets an increase of velocity towards the light signal. Hence a blue shift will be observed due to the Doppler effect. This is a co-ordinate invariant phenomenon. But it cannot be explained as a result of the Doppler effect by the observer in the rocket, since the detector is permanently at rest relative to the emitter in this reference frame. The observer in the accelerated rocket experiences a gravitational field. Light moves downwards in this field of gravity. Hence, Einstein concluded that light is blue shifted when it moves downwards in a field of gravity, and red shifted when it moves upwards.

Einstein then described a stationary situation in which light waves move downwards in a gravitational field. A certain number of waves per second enter a room through the ceiling. Due to the gravitational frequency shift a larger number leaves through the floor per second. This seems to be impossible in a stationary situation. The resolution – said Einstein – is that each second is a little longer at the floor than at the ceiling. In this way he deduced the gravitational time shift: Time goes slower farther down in a gravitational field. For an observer staying far down in a gravitational field the effect is opposite. He will observe that the rate of time is faster higher up in the gravitational field.

Let us consider the twins again. B observes twin A and the Earth and Alpha Proxima move with a velocity *v = 0,8c* a Lorentz contracted distance

$$L = L_0 \sqrt{1 - v^2/c^2} = 2,4 \; light \; years \tag{3}$$

According to B the time taken by A's travel to Alpha Proxima and back is

$$t_B(B) = 2L/v = 6 \, years \tag{4}$$

i.e. B predicts that he ages by 6 years during A's travel. This is in agreement with A's prediction,

$$t_B(B) = t_A(B) \tag{5}$$

But due to the kinematical time dilation B would predict that A ages by

$$t_B(A)_{OUT-IN} = \sqrt{1-v^2/c^2} \, t_B(B) = \sqrt{1-v^2/c^2} \, \frac{2\sqrt{1-v^2/c^2} L_0}{v} = \left(1 - \frac{v^2}{c^2}\right) \frac{2L_0}{v} = 3,6 \, years \tag{6}$$

which is in conflict with A's prediction that he should age by ten years, $T_A(A) = 10 \, years$.

Our suspicion is that something is missing in B's prediction of A's ageing as given above. Let us take a closer view upon what happens with A, according to B, when B turns at Alpha Proxima. When B accelerates towards his brother he experiences a field of gravity away from A, who is higher up in this gravitational field than he is. Hence as measured by B twin A at the Earth ages faster than B during the time,

$$\Delta t_B(B) = 2v/g \tag{7}$$

when B accelerates.

If B has constant proper acceleration it follows from the general theory of relativity that the relation between A's ageing and B's is [4]

$$\Delta t_B(A) = \left(1 + \frac{gL_0}{c^2}\right) \Delta t_B(B) \tag{8}$$

which gives

$$\Delta t_B(A) = \left(1 + \frac{gL_0}{c^2}\right) \frac{2v}{g} = \frac{2v}{g} + \frac{2vL_0}{c^2} \tag{9}$$

When the Earthbound twin A calculated his own and B's ageing during B's travel he neglected the time taken by A at Alpha Proxima to reverse his velocity. This means that his calculation is correct only in the limit of an infinitely large acceleration, $g \to \infty$. The expression for the ageing of A as calculated by B during the time B experiences a gravitational field, then reduces to

$$\Delta t_B(A) = \frac{2vL_0}{c^2} = 2 \cdot 0,8 \cdot 4 \, years = 6,4 \, years \tag{10}$$

Hence, the total ageing of A as correctly predicted by B is

$$t_B(A) = t_B(A)_{OUT-IN} + \Delta t_B(A) = \left(1 - \frac{v^2}{c^2}\right)\frac{2L_0}{v} + \frac{2vL_0}{c^2} = \frac{2L_0}{v} = 10 \text{ years} \quad (11)$$

in agreement with A's own prediction.

We have considered the twin paradox in flat spacetime and the conclusion seems to be: The twin who accelerates when they are away from each other is youngest when they meet after the travel [5]. Eriksen and Grøn [6] have considered the more general situation when both twins accelerate and found that the twin who has greatest acceleration is youngest after the travel. One may wonder if this is valid in general in a curved spacetime.

Also there is a problem with the principle of relativity. In the treatment above we have only decided in beforehand that A is not accelerated and B is, and then shown how we can calculate A's and B's ageing both from A's and B's point of view when each consider themselves as at rest. But how can we be sure that B is younger than A when they meet after the separation? Implicitly we have treated acceleration as absolute, and said that B has been accelerated and A not. At least this is so if space is globally empty except for the twins. Then there is a problem in that the gravitational field experienced by B is without any cause. If the principle of relativity breaks down for accelerated motion, B cannot consider himself as at rest. One may also wonder if curved spacetime can cure the problem with absolute acceleration in a globally empty, flat spacetime.

### 3. Ageing in curved spacetime

How should a twin move to age as fast as possible? In this connection J. Dorling [7] writes that every body has a privileged set of states of motion, namely those where it is moving along a path of maximal proper time, i.e. a geodesic curve. In so far as its trajectory deviates from such a path, non-gravitational forces must be invoked. Hence, the answer seems to be that the twin who moves along a geodesic curve between two events in spacetime ages fastest, i.e. that a freely moving twin ages fastest.

M. A. Abramowitz and S. Bajtlik [8] have, however, shown that there exist situations where the proper time is *not* maximal along a geodesic curve. They considered two twins in the Schwarzschild spacetime, A at rest and B circling the Earth, and found that the ratio of their proper travelling times is

$$\frac{\tau_A}{\tau_B} = \sqrt{\frac{1 - v_A^2/c^2}{1 - v_B^2/c^2}} \quad (12)$$

Hence, the twin A with $v_A = 0$ is oldest after the travel,

$$\tau_A = \frac{\tau_B}{\sqrt{1 - v_B^2/c^2}} \quad (13)$$

### 3.1. The concept acceleration in the theory of relativity

There are two quantities called "acceleration": Three-acceleration and four-acceleration. Three-acceleration is defined as the derivative of the coordinate velocity with respect to coordinate time. It is a relative acceleration which can be transformed away.

Four-acceleration is defined as the derivative of the four-velocity with respect to proper time. It is an absolute acceleration which cannot be transformed away. Four-acceleration is the acceleration of a particle as measured in an instantaneous inertial rest frame of the particle. Particles falling freely have vanishing four-acceleration. A non-vanishing four-acceleration is due to non-gravitational forces.

The heading of Abramowitz and Bajtlik's preprint is: "Adding to the paradox: the accelerated twin is older". The reason for this heading is their result cited above. They write: Twin A is accelerated and the twin B is not. From eq.(13) follows that at reunion the accelerated twin is older than his non-accelerated brother! Clearly by accelerated they mean that twin A is not in free fall, so he has a non-vanishing four-acceleration.

It should be noted, however, that A has no three-acceleration while B has a non-vanishing centripetal acceleration. Hence, like in Minkowski spacetime, the twin with vanishing three-acceleration is older.

In the "standard resolution" of the twin paradox presented above one considers two twins in Minkowski spacetime, and "acceleration" is usually meant to be a three-acceleration. However, in this case there is a degeneracy between three-acceleration and four-acceleration. For twins in flat spacetime the *invariant* statement would be: The twin with a non-vanishing four-acceleration is younger. In other words: the freely falling twin ages fastest. It is this statement that Abramowicz and Bajtlik have shown is not generally true in curved spacetime.

One may wonder: What *is* generally true? Abramowitz and Bajtlik asked: "Could the notion "the twin who moves faster, is younger at the reunion" be somehow extended to the classical version of the paradox in the Minkowski spacetime, for example by referring to the starry sky above the twins?". They left this question unanswered. The question has been investigated by S. Braeck and Ø. Grøn [9] by considering several versions of the twin paradox.

### 4. Twins with vertical motion

### 4.1. Vertical motion in a uniformly accelerated reference frame

Twin A stays at rest in the uniformly accelerated reference frame. Twin B is shot upwards and is then falling freely in a uniform gravitational field. Then the relationship between A's and B's ageing while they were away from each other is [9]

$$\frac{g\tau_B}{2c} = \sinh\left(\frac{g\tau_A}{2c}\right) \qquad (14)$$

Since $\sinh x > x$ it follows that $\tau_B > \tau_A$. In other words, the travelling twin (twin B) is older than the twin who stays at rest (twin A) at the reunion.

The situation described here is similar to the situation discussed by Abramowicz and Bajtlik. Twin A, who is at rest, has vanishing three-acceleration, but is not freely falling, and therefore has a non-vanishing four-acceleration. Twin B is traveling and has a non-vanishing three-acceleration, but he is freely falling and has a vanishing four-acceleration. Hence, in this case the twin who has a non-vanishing three-acceleration, vanishing four-acceleration and moves faster is older at the reunion, in contrast to what was found in the example presented by Abramowicz and Bajtlik. We conclude, therefore, that four-acceleration, three-acceleration and velocity cannot be decisive factors in determining which twin becomes the older.

### 4.2 Vertical and circular motion in the Schwarzschild spacetime

The same situation is now considered in the Schwarzschild spacetime. The twin A is at rest and B moves freely in the vertical direction. Again the result is that twin A is younger at the reunion [9]. These calculations thus demonstrate that vertical motion in Schwarzschild spacetime gives opposite result to that with circular motion considered by Abramowicz and Bajtlik.

We then have the following possibility in the Schwarzschild spacetime. Three triplets meet at an event P. One, A, remains at rest, the second, B, is shot upwards, moves freely and falls down again, and the third, C, moves freely along a circular path. They arrange the motions so that they meet again at an event Q. Then they compare their ageing, i.e. their increase of proper time between the events P and Q. The general theory of relativity predicts the following result: $\tau_B > \tau_A > \tau_C$, i.e. the triplet that moved along the circular geodesic path is youngest, and the triplet that moved along the vertical geodesic path is oldest. Timelike geodesics have in general extremal proper time between two events, but the proper time along a geodesic curve can be either maximal or minimal.

### 5. A cosmic time effect

We now consider the situation with A at rest and B in circular motion in the Schwarzschild spacetime from the point of view of a rotating reference frame in which twin B is at rest. A set of comoving coordinates $t'$, $r'$, $\theta'$, $\phi'$ in the rotating reference frame, is given by the transformation

$$t'=t , \quad r'=r , \quad \theta'=\theta , \quad \phi'=\phi-\omega t \tag{15}$$

Here $\omega > 0$ represents the angular velocity of the reference frame. Note that the coordinate clocks showing $t'$ goes at the same rate independent of their distance from the origin. For simplicity we assume that the two twins perform orbital motion at a constant radius $r_0$ in the equatorial plane for which $\theta' = \pi/2$. Then the line element in the rotating reference frame along the path of the twins takes the form

$$ds^2 = -\left(1-\frac{R_S}{r_0}-\frac{r_0^2\omega^2}{c^2}\right)c^2 dt'^2 + r_0^2 d\phi'^2 + 2r_0^2 \omega\, d\phi\, dt \tag{16}$$

For timelike intervals the general physical interpretation of the line element is that it represents the proper time $d\tau$ between the events connected by the interval,

$$ds^2 = -c^2 d\tau^2 \tag{17}$$

It follows that the proper travelling time measured by twin A's clock is

$$\tau_A = \left(1-\frac{R_S}{r_0}-\frac{r_0^2\omega^2}{c^2}-\frac{r_0^2\Omega^2}{c^2}-\frac{2r_0^2\omega\Omega}{c^2}\right)^{1/2} \Delta t'$$

(18)

where $\Omega = d\phi'/dt'$ is the angular velocity of the twin A in the rotating reference frame. The travelling time of B, having $\Omega_B = 0$ is

$$\tau_B = \left(1 - \frac{R_S}{r_0} - \frac{r_0^2 \omega^2}{c^2}\right)^{1/2} \Delta t' \qquad (19)$$

The terms in eq.(18) have the following physical interpretations:
$R_S/r_0$ represents the gravitational time dilation due to the central mass.
$r_0^2 \omega^2 / c^2$ represents the gravitational time dilation due to the centrifugal gravitational field.
$r_0^2 \Omega^2 / c^2$ represents the kinematical, velocity dependent time dilation for clocks moving in the rotating frame.
$2r_0^2 \omega \Omega / c^2$ is neither a gravitational nor a kinematical time dilation. It has not earlier been given any reasonable interpretation. Bræck and Grøn [9] have called it a *cosmic time effe*ct for reasons that will be explained below.

The expression for A's travelling time may be written

$$\tau_A = \left(1 - \frac{R_S}{r_0} - \frac{r_0^2 \omega^2}{c^2} + \frac{r_0^2}{c^2} f(\Omega)\right)^{1/2} \Delta t' \qquad (20)$$

where $f(\Omega) = -\Omega^2 - 2\omega\Omega$. The graph of the function $f(\Omega)$ is shown in Figure 1,

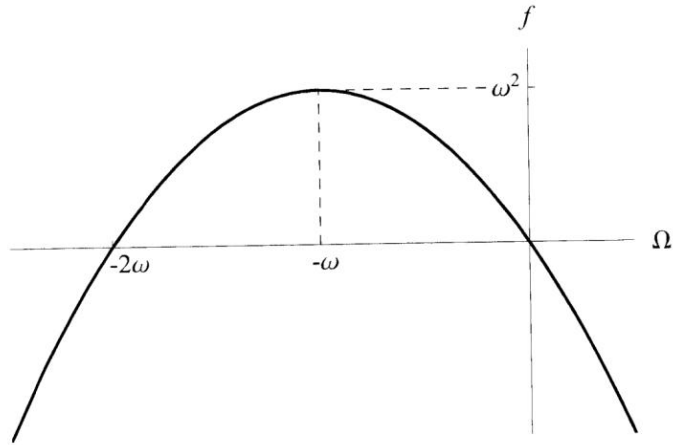

**Figure 1.** Sketch of the function $f(\Omega)$ introduced in Eq.(20) for different coordinate velocities $\Omega$.

The graph shows that a twin with $\Omega = -\omega$ ages fastest. This twin is at rest in the non-rotating inertial frame. Naturally the graph is symmetrical about this angular velocity. Hence for twins at the same height, the cosmic time effect acts so that the twin at rest in the non-moving, inertial frame ages fastest.

### 6. Ageing in the Kerr spacetime

The rotation of a mass distribution changes the properties of space outside it. Inertial frames are dragged along in the same direction as the mass rotates. We shall consider circular motion in an axially symmetric space. Along the circular path the line element can be written

$$ds^2 = g_{tt}dt^2 + 2g_{t\phi}dt\,d\phi + g_{\phi\phi}d\phi^2 \tag{21}$$

The coordinate clocks showing $t$ goe equally fast everywhere. Hence the proper time interval of a twin with angular velocity $\Omega = d\phi/dt$ is given by

$$d\tau = \left(-g_{tt} - 2g_{t\phi}\Omega - g_{\phi\phi}\Omega^2\right)^{1/2} dt \tag{22}$$

It can be shown [6] that an observer with zero angular momentum (ZAMO) has angular velocity

$$\Omega_Z = -g_{t\phi}/g_{\phi\phi} \tag{23}$$

A non-vanishing value of $\Omega_Z$ is an expression of inertial dragging. Let us find the angular velocity of the twin who ages fastest. One might think that it is the twin at rest due to the kinematical time dilation which tends to slow down the ageing. Putting the derivative of the function

$$F(\Omega) = -g_{tt} - 2g_{t\phi}\Omega - g_{\phi\phi}\Omega^2 \tag{24}$$

equal to zero, one finds, however, that the ZAMO ages fastest.

In the Kerr spacetime the angular velocity of a ZAMO is [4]

$$\Omega_Z = \frac{2mac}{r^3 + ra^2 + 2ma^2} \tag{25}$$

where $m = GM/c^2$ is the gravitational length of the central rotating body, and $a = J/Mc$ where $J$ is the angular momentum of the central mass (note that $a$ has dimension length). The ZAMO angular momentum vanishes in the asymptotic Minkowski spacetime in the limit $r \to \infty$. If the central body is non-rotating there is Schwarzschild spacetime and the angular velocity of the ZAMO vanishes.

Our treatment of the twins in the Schwarzschild and Kerr spacetimes seems to imply that rotating motion is absolute. For example one can decide which twin rotates by measuring how fast he ages. In the special theory of relativity rotational motion is absolute. However if the general principle of relativity is generally valid according to the general theory of relativity, rotational motion has to be relative. Whether this is so is still discussed. The phenomenon of perfect inertial dragging plays a decisive role in this connection.

### 7. Inertial dragging inside a rotating shell of matter

#### 7.1. The weak field result

Inertial dragging inside a rotating shell of matter was described already in 1918 by H. Thirring [10]. He calculated the angular velocity of $\Omega_Z$ a ZAMO inside a shell with Schwarzschild radius $R_S$ and radius $r_0$ rotating slowly with angular velocity $\omega$, in the weak field approximation and found the inertial dragging angular velocity,

$$\Omega_Z = \frac{8R_S}{3r_0}\omega \qquad (26)$$

This calculation does not, however, remove the difficulty with absolute rotation in an asymptotically empty Minkowski space. Both the angular velocity of the shell and that of the ZAMO are defined with respect to a system that is non-rotating in the far away region. There is nothing that determines this system. The absolute character of rotational motion associated with the asymptotically empty Minkowski spacetime, has appeared.

### 7.2. Perfect inertial dragging

In 1966 D. R. Brill and J. M. Cohen [11] presented a calculation of the ZAMO angular velocity inside a rotating shell valid for arbitrarily strong gravitational fields, but still restricted to slow rotation, giving

$$\Omega_Z = \frac{4R_S}{r_0 + R_S}\frac{2r_0 - R_S}{3r_0 - R_S}\omega \qquad (27)$$

For weak fields, i.e. for $r_0 \gg R_S$, this expression reduces to that of Thirring. But if the shell has a radius equal to its own Schwarzschild radius, $r_0 = R_S$, the expression above gives $\Omega_Z = \omega$. Then there is *perfect dragging*. In this case the inertial properties of space inside the shell no longer depend on the properties of the ZAMO at infinity, but are completely determined by the shell itself.

Brill and Cohen further write that a shell of matter with radius equal to its Schwarzschild radius together with the space inside it can be taken as an idealized cosmological model, and proceeds: "Our result shows that in such a model there cannot be a rotation of the local inertial frame in the center relative to the large masses in the universe. In this sense our result explains why the "fixed stars" are indeed fixed in our inertial frame.

In 1995 H. Pfister [12] wrote that whether there exists an exact solution of Einstein's field equations with flat spacetime and correct expressions for the centrifugal- and Coriolis acceleration inside a rotating shell of matter, was still not known. However, permitting singular shells such a solution certainly exists, as will now be made clear.

### 7.3. A source of the Kerr metric with perfect inertial dragging

In 1981 C. A. Lopez [13] found a source of the Kerr spacetime. A few years later Ø. Grøn [14] gave a much simpler deduction of this source and discussed some of its physical properties. The source is a shell with radius $r_0$ rotating with an angular velocity

$$\omega = \frac{ac}{a^2 + r_0^2} \qquad (28)$$

The radius of the exterior horizon in the Kerr metric is

$$r_+ = m + \sqrt{m^2 - a^2} \qquad (29)$$

Hence, if the radius of the shell is equal to the horizon radius $r_0 = r_+$, the ZAMO angular velocity just outside the shell is equal to the angular velocity of the shell,

$$\Omega_Z(r_+) = \omega(r_+) = \frac{ac}{2mr_+} \qquad (30)$$

Demanding continuity of the dragging angular velocity at the shell it follows that the inertial frames in the Minkowski spacetime inside the shell are co-moving with the shell. There is perfect dragging of the inertial frames inside the shell. The properties of the shell, and of spacetime outside and inside the shell, solve Einstein's field equations without needing the assumptions of week fields and slow rotation. The inertial properties of space inside the shell, such as the Coriolis acceleration, do not depend on any property of an asymptotic far away region, only on the state of motion of the reference frame relative to the shell.

### 8. Is there perfect dragging in our universe?

The distance that light and the effect of gravity have moved since the Big Bang is called the lookback distance, $R_0 = ct_0$, where $t_0$ is the age of the universe. WMAP-measurements have shown that the age of the ΛCDM-model of our universe is close to its Hubble-age, $t_H = 1/H_0$, namely that $t_0 = 0,996 t_H$, and that the universe is flat, i.e. that it has critical density

$$\rho_{cr} = 3H_0^2 / 8\pi G \qquad (31)$$

It follows that

$$8\pi G \rho_{cr} / 3c^2 = (H_0/c)^2 \approx 1/R_0^2 \qquad (32)$$

The Schwarzschild radius of the cosmic mass inside the lookback distance is

$$R_S = 2GM/c^2 = (8\pi G \rho_{cr}/3c^2) R_0^3 \approx R_0 \qquad (33)$$

Hence in our universe the Schwarzschild radius of the mass within the lookback distance is approximately equal to the lookback distance. It follows that the condition for perfect dragging may be fulfilled in our universe.

The question of perfect dragging in our universe has been considered from a different point of view by C. Schmid [15, 16]. By introducing a rotational perturbation in a realistic FRW-model he has shown that the ZAMO angular velocity in the perturbed FRW universe is equal to the average angular velocity of the cosmic mass distribution. Hence perfect dragging explains why the swinging plane of the Foucault pendulum rotates with the "starry sky". In Newtonian gravity where there is no dragging, this is a consequence of the absolute character of rotation. One says that the swinging plane of the Foucault pendulum is at rest relative to the starry sky because neither of them rotates. Hence the pendulum is in a room with an absolute rotation.

### 9. An extended model of Minkowski spacetime

Accelerated and rotational motion is absolute according to the general theory of relativity if the asymptotically far away regions in the Schwarzschild- and Kerr spacetimes are imagined to be globally empty Minkowski spacetime. Then the general principle of relativity is not valid. The previous results show, however, that the general principle of relativity can be saved by introducing an extended model of Minkowski spacetime in which space is completed by a far away cosmic shell of mass with radius equal to the horizon radius of the space outside the shell. This radius may be set equal to the lookback distance of the universe. The shell then represents the cosmic mass inside the lookback distance, i.e. the mass that may act causally upon us.

The extended model of Minkowski spacetime is also relevant in connection with a point made several years ago by C. Møller [17]. He wrote that when one solves Einstein's field equations in a rotating reference frame it is necessary to take account of the far away cosmic masses. However there was an exception for globally or asymptotic Minkowski spacetime, where there was no cosmic masses. In the extended model the Minkowski spacetime is treated in the same way as any other spacetime.

In the spacetime inside the shell a centrifugal gravitational field appears in a reference frame rotating relative to the shell. An observer in a frame R rotating relative to the shell can maintain that the frame R does not rotate, and that it is the shell that rotates. His calculations would show that there is perfect dragging inside the rotating shell, and that this causes the centrifugal gravitational field. With this model of the Minkowski spacetime rotational motion is relative. Without the shell rotation is absolute.

Translational inertial dragging inside an accelerating shell has been investigated in the weak field approximation by Ø. Grøn and E. Eriksen [18]. They found that the inertial acceleration inside a shell with acceleration $g$, Scwarzschild radius $R_S$ and radius $R$ is

$$a = \frac{11}{6}\frac{R_S}{R}g \tag{34}$$

Hence, according to this approximate calculation there is perfect translational dragging inside a shell with radius $R = (11/6)R_S$.

### 10. Conclusion

The assumption that spacetime outside a central mass distribution has an asymptotically flat far away region which is globally empty is inconsistent with the general principle of relativity. In such a spacetime the accelerated twin cannot say that he is at rest because the gravitational field he experiences has no source. It is an ad hoc gravitational field introduced into the description when we say that twin A is at rest and B travels. In order that both twins shall have the right to claim that they are at rest, we have to introduce the extended model of the Minkowski spacetime. Then the field of gravity experienced by B is due to perfect dragging induced by the accelerating cosmic shell.

Consider a non-rotating central mass inside the extended Minkowski spacetime and a twin moving around it. In this space a ZAMO is not moving relative to the cosmic shell. Among all twins moving along the same circular path, the ZAMO ages fastest. The cosmic shell determines not only the inertial properties of spacetime inside it, but also its temporal properties. This is the physical significance of the cosmic time effect.

**References**
1. A. Einstein, "The Foundations of the General Theory of Relativity", translated from "Die Grundlage der allgemeinen Relativitätstheorie", Annalen der Physik, **49** (1916), in *The Principle of Relativity*, 111-164, Dover Publications (1952).
2. L. Iorio, "An analytical treatment of the Clock Paradox in the framework of the special and general theories of relativity". Found. Phys. Lett. **18**, 1-19 (2006).
3. E. Minguzzi, "Differential ageing from acceleration, an explicit formula". Am. J. Phys. **73**, 876-880 (2005).
4. Ø. Grøn. "Lecture Notes on the General Theory of Relativity", p.90. Springer (2009).
5. Ø. Grøn. "The twin paradox in the theory of relativity", Eur. J. Phys. **27**, 885-889 (2006).
6. E. Eriksen and Ø. Grøn. "Relativistic dynamics in uniformly accelerated reference frames with application to the clock paradox", Eur. J. Phys. **39**, 39-44, (1990).
7. J. Dorling. "Did Einstein need General Relativity to solve the Problem of Absolute Space? Or had the Problem already been solved by Special Relativity? Brit. J. Philos. Sci. **29**, 311-323 (1978).
8. M. A. Abramowitz and S. Bajtlik. "Adding to the paradox: the accelerated twin is older", ArXiv: 0905.2428 (2009).
9. S. Braeck and Ø. Grøn. ArXiv: 0909.5364 (2009).
10. H. Thirring. "Über die Wirkung rotierende ferner Massen in der Einsteinschen Gravitationstheorie", Physikalische Zeitschrift **19**, 33-39 (1918).
11. D. R. Brill and J. M. Cohen. "Rotating masses and their effect on inertial frames", Phys. Rev. **143**, 1011-1015 (1966).
12. H. Pfister. "Dragging effects near rotating bodies and in Cosmological Models", in *Mach's Principle.* Ed. J. Barbour and H. Pfister. Birkhauser (1995).
13. C. A. Lopez. "Extended model of the electron in general relativity", Phys. Rev. D **30**, 313-316 (1984).
14. Ø. Grøn. "New derivation of Lopez's source of the Kerr-Newman field", Phys. Rev. D **32**, 1588-1589 (1985).
15. C. Schmid. "Cosmological gravitomagnetism and Mach's principle", Phys. Rev. D **74**, 044031 (2006).
16. C. Schmid. "Mach's principle: Exact frame-dragging via gravitomagnetism in perturbed Friedmann-Robertson-Walker universes with k=(±1, 0)", Phys. Rev. D **79**, 064007 (2009).
17. C. Møller. "The Theory of Relativity". Chap. 8. Oxford (1952).
18. Ø.Grøn and E. Eriksen. "Translational inertial dragging", Gen. Rel. Grav. **21**, 105-124 (1989).